\documentclass{article}

\usepackage{graphicx}

\def\abstract#1{\vskip 7mm 
        \begin{center}{\large Abstract}\par \smallskip
                \begin{minipage}[c]{12cm}
                        \small #1
                \end{minipage}
        \end{center}
}
\def\title#1{\begin{center}{\Large\bf #1}\end{center}}
\def\author#1{\vskip 5mm \begin{center}{#1}\end{center}}
\def\address#1{\begin{center}{\it #1}\end{center}}
\makeatletter
\@ifundefined{lesssim}{}{}
\@ifundefined{gtrsim}{}{}
\def\vereq#1#2{\lower3pt\vbox{\baselineskip1.5pt \lineskip1.5pt
\ialign{$\m@th#1\hfill##\hfil$\crcr#2\crcr\sim\crcr}}}
\makeatother

\font\sm=cmr8
\def\black{}
\def\red{}

\def\blue{}

\def\bs{\bigskip}
\def\ms{\medskip}

\def\ni{\noindent}

\begin{document}

\title{Angular momentum and conservation laws for dynamical black holes}
\author{\blue Sean A. Hayward\black\footnote{E-mail:sean\_a\_hayward@yahoo.co.uk}}
\address{Institute for Gravitational Physics and Geometry, The Pennsylvania
State University, University Park, PA 16802, U.S.A.}

\abstract{Black holes can be practically located (e.g.\ in numerical 
simulations) by \blue trapping horizons\black{}, hypersurfaces foliated by 
marginal surfaces, and one desires physically sound measures of their \blue 
mass \black and \blue angular momentum\black{}. A generically unique angular 
momentum can be obtained from the Komar integral by demanding that it satisfy 
a simple \blue conservation law\black{}. With the irreducible (Hawking) mass 
as the measure of energy, the conservation laws of energy and angular momentum 
take a similar form, expressing the rate of change of mass and angular 
momentum of a black hole in terms of fluxes of energy and angular momentum, 
obtained from the matter energy tensor and an \blue effective energy tensor 
for gravitational radiation\black{}. Adding charge conservation for 
generality, one can use Kerr-Newman formulas to define combined energy, 
surface gravity, angular speed and electric potential, and derive a dynamical 
version of the so-called ``\blue{}first law\black{}'' for black holes. A 
generalization of the ``\blue{}zeroth law\black{}'' to local equilibrium 
follows. Combined with an existing version of the ``\blue{}second 
law\black{}'', all the key quantities and laws of the classical paradigm for 
black holes (in terms of Killing or event horizons) have now been formulated 
coherently in a general dynamical paradigm in terms of trapping horizons.}

\section{Komar integral and twist: $J[\psi]$, $\omega$}

\vskip-1cm\hfill\includegraphics[height=35mm]{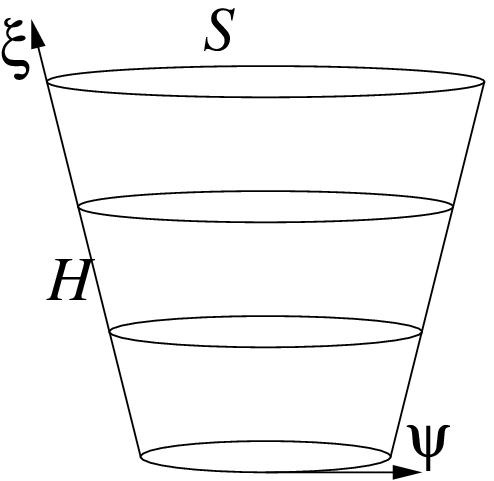}

\vskip-25mm\ni A 1-parameter family of topologically spherical spatial 
surfaces \red$S\black$ 

locally forms a foliated hypersurface \red$H\black$.

\ni A \blue generating vector \red$\xi^a\black=(\partial/\partial x)^a$ 
generates the constant-$x$ surfaces $S$, 

and can be taken to be \blue normal\black{}, $h_{ab}\xi^b=0$, 

where \red$h_{ab}\black$ is the induced metric of $S$. 

\ni The Komar integral is {\sm [Komar 1959]} 
\red$$J[\psi]\black=-{1\over{16\pi}}\oint_S{*}\epsilon_{ab}\nabla^a\psi^b,$$

where $\epsilon_{ab}$ is the antisymmetric 2-form of the normal space 

and \red${*}1\black=\sqrt{\det h}\,d\theta\wedge d\phi$ is the \blue area form 
\black of $S$. 

\ni One can take null coordinates \red$x^\pm\black$ for the normal space, 

labelling the outgoing and ingoing null hypersurfaces passing through each 
$S$. 

\ni Then $\epsilon_{ab}=e^{2\varphi}(dx^+_adx^-_b-dx^-_adx^+_b)$ in terms of a 
normalization function \red$e^{-2\varphi}\black=-g^{ab}dx^+_adx^-_b$. 

\ni For a \blue transverse \black vector $\psi^a$, $h^a_b\psi^b=\psi^a$, the 
Komar integral can be rewritten as 
\blue$$J[\psi]={1\over{8\pi}}\oint_S{*}\psi^a\omega_a\black$$

where 
\red$\omega_c\black=\frac12e^{2\varphi}h_c^b(dx^+_a\nabla^adx^-_b-dx^-_a\nabla^adx^+_b) 
=\frac12e^{2\varphi}h_{bc}[dx^+,dx^-]^b$ is the \blue twist\black{}, 

measuring the non-integrability of the normal space {\sm [Hayward 1993]}. 

\ni The twist is invariant under relabelling $x^\pm\to\tilde x^\pm(x^\pm)$ and 
therefore is an invariant of $H$ 

unless $\xi^a$ becomes null, so the twist expression for $J[\psi]$ is also an 
invariant of $H$. 

\ni The gauge dependence of the Komar integral for a single $S$ is fixed by 
$H$.

\section{Uniqueness: $\psi$}

\ni Assume that the axial vector $\psi^a$ has vanishing transverse divergence, 
\blue$D_a\psi^a=0\black$,

where \red$D_a\black$ is the covariant derivation of $h_{ab}$.

\ni Then $J[\psi]$ can be defined equivalently in terms of other normal 
fundamental forms 

differing by a gradient, $\omega_a\mapsto\omega_a+D_a\lambda$. 

\ni There are several such expressions, though they are gauge-dependent, 
fixing $\varphi=0$

{\sm [Brown \& York 1993, Ashtekar, Beetle \& Lewandowski 2001, Ashtekar \& 
Krishnan 2002, Booth \& Fairhurst 2004]}.

\ms\ni To obtain a conservation law for angular momentum, expressing 
\red$L_\xi\black J[\psi]$ (Lie derivative), 

it is natural to propagate $\psi^a$ along $\xi^a$ by 
\blue$L_\xi\psi^a=0\black$ {\sm [Gourgoulhon 2005]}. 

\ni There is a commutator identity $L_\xi(D_a\psi^a)-D_a(L_\xi\psi^a)=\psi^a 
D_a\theta_\xi$ for any normal vector $\xi^a$ 

and transverse vector $\psi^a$, where \red$\theta_\xi\black$ is the \blue 
expansion \black along $\xi^a$, ${*}\theta_\xi=L_\xi({*}1)$.

\ni So \blue$\psi^a D_a\theta_\xi=0\black$.

This is automatic if $D_a\theta_\xi=0$, as in spherical symmetry or along a 
null trapping horizon. 

\ni However, for generic $H$, one expects $D_a\theta_\xi\not=0$ almost 
everywhere. 

\ni The hairy ball theorem states that a continuous vector field 
($D^a\theta_\xi$) 

must vanish somewhere on a sphere; however,

a generic situation is that the curves \red$\gamma\black\subset S$ of constant 
$\theta_\xi$ 

form a smooth foliation of circles with two poles. 

\vskip-2cm\hfill\includegraphics[height=3cm]{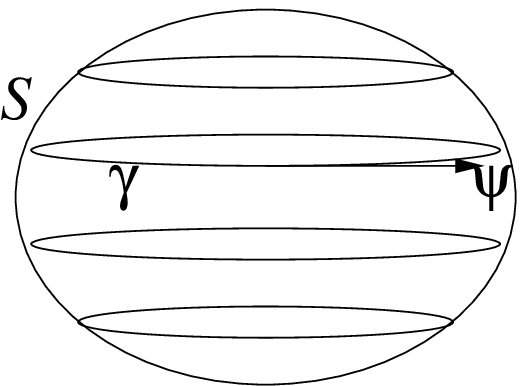}

\vskip-1cm\ni Assuming so, $\psi^a$ must be tangent to $\gamma$. 

\ni Then one can find a \blue unique \black $\psi^a$, up to sign, in terms of 
the unit tangent vector \red$\hat\psi^a\black$ 

and arc length \red$ds\black$ along $\gamma$: 
\blue$\psi^a=\hat\psi^a\oint_\gamma ds/2\pi\black$.
 
\ni Then the angular momentum becomes unique up to sign, $J[\psi]=\red 
J\black$. 

The sign is naturally fixed by $J>0$ (if $J\not=0$) and continuity of $\psi^a$.

\ms\ni For an axisymmetric space-time with axial Killing vector $\psi^a$, one 
has $D_a\psi^a=0$. 

\ni Assuming that $\xi^a$ respects the symmetry, $0=L_\psi\xi^a=-L_\xi\psi^a$, 

so the above construction, if unique, yields the correct $\psi^a$. 

\ni For example, consider a Kerr space-time in Boyer-Lindquist coordinates 
$(t,r,\theta,\phi)$, 

with $S$ given by constant $(t,r)$ and $\xi^a=(\partial/\partial r)^a$. 

\ni If $ma\not=0$, $D_a\theta_\xi$ is a certain function of $\theta$ (and 
$r$), 

non-zero except at the poles and equator (and isolated values of $r$), 

so that a unique continuous $\psi^a$ exists, 
$\psi^a=(\partial/\partial\phi)^a$.

\section{Conservation: $\Theta$}

\vskip-1cm\hfill\includegraphics[height=2cm]{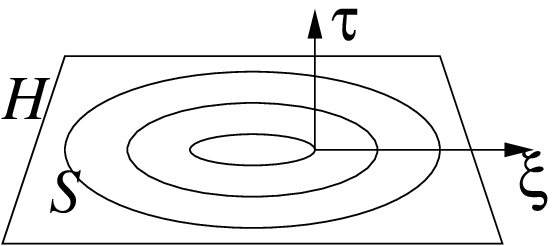}

\vskip-1cm\ni Introduce the normal vector \red$\tau^a\black$ dual to $\xi^a$: 

$h_{ab}\tau^b=0$, $g_{ab}\tau^a\xi^b=0$, 
$g_{ab}\tau^a\tau^b=-g_{ab}\xi^a\xi^b$. 

\ni Its expansion \red$\theta_\tau\black$ is given by 
${*}\theta_\tau=L_\tau({*}1)$.

\ni There is a simple expression for the rate of change of angular momentum if 
\blue$\psi^aD_a\theta_\tau=0\black$, 

which is generally inconsistent with the above constraint $\psi^a 
D_a\theta_\xi=0$. 

\ni However, it is consistent: 

(i) along a null $H$, $\tau^a=\xi^a$ {\sm [Damour 1978]}; 

(ii) along a \blue trapping horizon \black$H$, $|\theta_\tau|=|\theta_\xi|$ 
{\sm [Ashtekar \& Krishnan, Booth \& Fairhurst, Gourgoulhon]}; 

(iii) along \blue uniformly expanding flows\black{}, 
$D_a\theta_\xi=D_a\theta_\tau=0$ {\sm [Hayward 1994]}. 

\ni The expression is 
$$L_\xi J=-\oint_S{*}\left(T_{ia}-\frac{h^{jk}D_k\sigma_{aij}}{16\pi}\right)\psi^i\tau^a$$
 
where \red$T_{ab}\black$ is the matter energy tensor, so that 
$T_{ia}\psi^i\tau^a$ is an angular momentum density,

and \red$\sigma_{aij}\black$ is the \blue shear \black form, the traceless 
part of the second fundamental form of $S$,

$\sigma_{\pm ij}=h_i^kh_j^lL_\pm h_{kl}-\frac12h_{ij}h^{kl}L_\pm h_{kl}$, 
where $L_\pm$ are Lie derivatives along the null normals. 

\ni Then one can identify the transverse-normal block 
\blue$$\Theta_{i\pm}=-{1\over{16\pi}}h^{jk}D_k\sigma_{\pm ij}\black$$ 

of an \blue effective energy tensor \red$\Theta_{ab}\black$ for gravitational 
radiation.

\ni The normal-normal block ($\Theta_{\pm\pm}$, $\Theta_{\pm\mp}$) occurs in 
the energy conservation law {\sm [Hayward 2004]},

with energy densities $\Theta_{\pm\pm}=||\sigma_\pm||^2/32\pi$ of ingoing and 
outgoing gravitational radiation,

recovering the Bondi energy flux at null infinity 

and the Isaacson energy density for high-frequency linearized gravitational 
waves.

\ni It seems that gravitational radiation is encoded in null shear 
$\sigma_{\pm ij}$, 

and that differential shear has angular momentum density $\Theta_{i\pm}\psi^i$.

\ms\ni Then \blue conservation of angular momentum \black takes the same form 
\blue 
$$L_\xi J=-\oint_S{*}(T_{ab}+\Theta_{ab})\psi^a\tau^b\black$$ 

as \blue conservation of energy \black{\sm [Hayward 2004]} \blue$$L_\xi 
M=\oint_S{*}(T_{ab}+\Theta_{ab})k^a\tau^b\black$$ 

for the Hawking mass \red$M\black$ along a trapping horizon or a uniformly 
expanding flow, 

where \red$k^a\black$ is the normal dual of $\nabla^aR$, \blue area 
\red$A\black=\oint_S{*}1$ defining \red$R\black$ by \blue$A=4\pi R^2\black$.

\section{Averagely conserved currents and charges: $j_{\{M,J,Q\}}$}

\ni For an electromagnetic field, \blue charge \red$Q\black$ and 
charge-current density \red$j_Q\black$ are related by 
$$[Q]=-\int_H{*}(j_Q^a\tau_a)\wedge dx=-\int_H\hat{*}j_Q^a\hat\tau_a$$

where the first expression holds for $H$ of any signature and the second for 
spatial $H$,

$\hat{*}1$ being the proper volume element and $\hat\tau^a$ the unit normal 
vector. 

\ni The surface-integral form is \blue$$L_\xi 
Q=-\oint_S{*}j_Q^a\tau_a\black{}.$$ 

\ni The above conservation laws can be written in the same form  \blue$$L_\xi 
M=-\oint_S{*}j_M^a\tau_a\black{},\quad\blue L_\xi 
J=-\oint_S{*}j_J^a\tau_a\black$$ 

by defining
$$(\red j_M\black)^a=-(T^{ab}+\Theta^{ab})k_b,\quad
(\red j_J\black)^a=(T^{ab}+\Theta^{ab})\psi_b.$$

\ni The physical interpretation of the components is 
 
$j_M=(\hbox{energy density},\hbox{energy flux})$,

$j_J=(\hbox{angular momentum density},\hbox{angular stress})$. 

$j_Q=(\hbox{charge density},\hbox{current density})$,

\ni For spatial $\xi$, $\oint_S{*}(j_M,j_J,j_Q)^a\xi_a 
=(\hbox{power},\hbox{torque},\hbox{current})$,

$-(j_M,j_J,j_Q)^a\tau_a=(\hbox{energy density},\hbox{angular momentum 
density},\hbox{charge density})$. 

\ni Local charge conservation takes the form \blue$\nabla_aj_Q^a=0\black$.

\ni For energy and angular momentum, one has only \blue quasi-local \black 
conservation laws: \blue$$\oint_S{*}\nabla_aj_M^a=\oint_S{*}\nabla_a 
j_J^a=0\black{}.$$

Then $j_M$ and $j_J$ are \blue averagely conserved\black{}.

\section{Laws of black-hole dynamics: $E$, $(\kappa,\Omega,\Phi)$}

\ni There are now three conserved quantities $(M,J,Q)$, as for a Kerr-Newman 
black hole. 

\ni One can use the Kerr-Newman formula for the ADM energy to define an energy 
\red$$E\black={\sqrt{((2M)^2+Q^2)^2+(2J)^2}\over{4M}}$$ 

for each marginal surface in a trapping horizon, where \blue$R=2M\black$. 

\ni Then \blue surface gravity 
\red$$\kappa\black={(2M)^4-(2J)^2-Q^4\over{2(2M)^3\sqrt{((2M)^2+Q^2)^2+(2J)^2}}},$$ 

\blue angular speed 
\red$$\Omega\black={J\over{M\sqrt{((2M)^2+Q^2)^2+(2J)^2}}}$$ 

and \blue electric potential 
\red$$\Phi\black={((2M)^2+Q^2)Q\over{2M\sqrt{((2M)^2+Q^2)^2+(2J)^2}}}$$ 

can be defined by thermostatic-style formulas 
$$\kappa=8\pi{\partial E\over{\partial A}}={1\over{4M}}{\partial E\over{\partial M}}, 
\quad\Omega={\partial E\over{\partial J}}, \quad\Phi={\partial E\over{\partial 
Q}}.$$ 

\ni There follows a dynamic version of the ``first law of black-hole 
mechanics'': \blue$$L_\xi E={\kappa\over{8\pi}}L_\xi A+\Omega L_\xi J+\Phi 
L_\xi Q\black,$$

really analogous to the Gibbs equation. 

\ni In energy-tensor form,

$$L_\xi 
E=\oint_S{*}\left((T_{ab}+\Theta_{ab})K^a\tau^b-\Phi j_Q^b\tau_b\right)$$

where \red$K^a\black=4M\kappa k^a-\Omega\psi^a$ reduces to the stationary 
Killing vector on a Kerr-Newman black hole.

\ms\ni For $J\ll M^2$ and $Q\ll M$, 
$$E\approx M+\textstyle{1\over2}I\Omega^2+\textstyle{1\over2}Q^2/R$$

where $J=I\Omega$ defines the \blue moment of inertia 
\red$$I\black=M\sqrt{((2M)^2+Q^2)^2+(2J)^2}=ER^2.$$ 

\ni Thus $E\ge M$ can be interpreted as a \blue combined energy\black{}, 
including the \blue irreducible mass \black$M$,

rotational kinetic energy $\approx{1\over2}I\Omega^2$ and electrostatic energy 
$\approx{1\over2}Q^2/R$. 

\ni Energy $E-M$ can be extracted by Penrose-type processes, while $L_\xi 
M\ge0$, 

assuming NEC, by the area law \blue$L_\xi A\ge0\black$ for black holes {\sm 
[Hayward 1994]}, cf.\ ``second law''. 

\ms\ni\blue Local equilibrium\black{}: 
$(j_M,j_J,j_Q)^a\tau_a=0\Rightarrow(M,J,Q)$ constant $\Rightarrow\blue\kappa$ 
constant\black{}, cf.\ ``zeroth law''.

\bs\ni Acknowledgements. Thanks to Abhay Ashtekar, Ivan Booth and Eric 
Gourgoulhon for discussions and to the conference organizers for local support 
and hospitality. Research supported by NSF grants PHY-0090091, PHY-0354932 and 
the Eberly research funds of Penn State.

\end{document}